# The rise of unconventional magnetism

Xiaobing Chen[1,2], Weizhao Chen[1,2] and Qihang Liu[1,2,*]

[1]*Quantum Functional Materials, Department of Physics, and Guangdong Basic Research Center of Excellence for Quantum Science, Southern University of Science and Technology (SUSTech), Shenzhen 518055, China*

[2]*Quantum Science Center of Guangdong–Hong Kong–Macao Greater Bay Area (Guangdong), Shenzhen 518045, China*

[*]Email: liuqh@sustech.edu.cn

**Abstract**

**Unconventional magnetism represents a paradigm shift in condensed matter physics, effectively bridging the fast, high-density advantages of antiferromagnets with the facile read-write capability of ferromagnets. Recent developments in spin space group theory have established a systematic methodology to decouple magnetic geometry from relativistic spin-orbit coupling, driving the exploration of unconventional magnets that exhibit compensated magnetization with time-reversal-odd responses. Here, we review unconventional magnetism across three pivotal facets in momentum space: spin textures, quantum geometry, and emergent quasiparticles. From the perspective of symmetry analysis, we elucidate the mechanisms underlying time-reversal-odd physical responses, including non-relativistic spin splitting, anomalous and nonlinear Hall effects, and exotic electronic and magnonic topological phases. Finally, we provide a forward-looking perspective on coupling unconventional magnetism with ferroelectricity, superconductivity, and moiré engineering. By exploiting symmetry-driven insights, this review highlights the functional potential of unconventional magnets in developing next-generation, high-speed, and energy-efficient spintronic devices.**

# I. Introduction

While conventional semiconductor architectures face significant challenges in device miniaturization and energy efficiency, spintronic technologies, including spin transfer torque and spin-orbit torque magnetic random-access memory, have emerged as potential solutions for high-speed and high-capacity data storage[1,2]. While traditional spintronic technologies are built upon the physics of ferromagnets (FMs), the net macroscopic magnetization of FMs introduce stray-field interference and relatively slow magnetization dynamics. Consequently, antiferromagnets (AFMs) have attracted growing interest as a platform for ultra-fast, high-density information processing by virtue of their compensated magnetic moments and terahertz-range dynamics[3-7].

Despite these advantages, the absence of a net magnetization in AFMs poses a significant challenge: how to efficiently read and write information in a system whose order parameter remains hidden to conventional magnetic probes[4,7]. Over the past decade, efforts to address this problem have relied primarily on relativistic spin-orbit coupling (SOC), enabling electrical control and detection of antiferromagnetic states[8-10]. However, the usage of SOC typically requires heavy metals and, more importantly, overshades the magnetic order's capability to directly govern electronic states through strong non-relativistic exchange interactions.

The difficulty in distinguishing magnetic geometry from SOC effects is deeply rooted in the historical framework of magnetic symmetry. Traditionally, the study of magnetic symmetry has been guided by two central propositions proposed by Bradley[11], which involve: 1) classifying magnetic orders within specific symmetry groups, and 2) identifying the physical properties dictated by these groups. However, the textbook framework of magnetic space group (MSG) assumes a complete locking between lattice and spin spaces[12]. This assumption naturally entangles the non-relativistic effects of magnetic geometry with the relativistic effects induced by SOC. Consequently, the MSG framework faces limitations in providing an exhaustive classification of magnetic phases[13,14]. To address this, the field has recently adopted the spin space group (SSG) framework[15-25]. By decoupling lattice and spin rotations, SSGs offer a precise description of magnetic geometry. This separation from SOC reveals that magnetic geometry alone can drive pronounced transport phenomena governed by the exchange interaction.

The clarity provided by this extended symmetry framework has identified unconventional magnetism[26] as a new research frontier. *These materials effectively bridge the gap between AFMs and FMs, combining the compensated magnetization of the former with various ferromagnetic-like, time-reversal-odd physical responses of the latter.* In this review, we first elucidate the fundamental connections and distinctions between SSGs and MSGs, the two frameworks describing magnetic geometry and SOC effects, respectively. We then systematically explore the landscape of unconventional magnetism through three pivotal facets in reciprocal space: spin textures, quantum geometry, and emergent quasiparticles (Fig. 1). Furthermore, we demonstrate that the interplay among these three facets gives rise to various time-reversal-odd signatures, including momentum-dependent spin splitting, the anomalous Hall effect, nonlinear quantum transport, and exotic electronic and magnonic topological states. Moreover, integrating additional degrees of freedom, such as superconductivity, ferroelectricity, and moiré engineering, provides pathways to couple with magnetic geometry, thereby enriching the functional phase space of these materials. By emphasizing the critical role of magnetic geometry, we aim to illustrate how these symmetry-driven insights are paving the way for a new generation of energy-efficient, high-performance spintronic devices.

**II. Symmetry frameworks: Spin group vs. magnetic group**

The physical properties of a crystal are largely governed by the symmetry of its Hamiltonian. Traditionally, it was believed that MSGs provided a complete framework for describing all magnetic phenomena. However, MSGs face limitations in distinguishing between ferromagnetic and antiferromagnetic orders and in precisely characterizing different magnetic propagation modes[27]. Furthermore, recent studies have revealed that electronic and transport properties in magnetic materials can arise purely from magnetic order, even in the absence of SOC[28-32]. This implies that such physical responses may originate from two distinct sources of symmetry breaking: the SOC effects and the magnetic order[16]. While the former is conventionally described by MSGs, the latter requires the more extensive framework of SSGs[16,19].

The evolution of crystallographic group theory reflects a progressive insight into the mechanisms by which distinct order parameters, such as the lattice, magnetic order, and SOC, induce successive stages of symmetry breaking. To systematically delineate different group frameworks, we consider the evolution of the single-particle electronic Hamiltonian[19] (Fig. 2). Beginning with a SOC-free, non-magnetic crystal described by $H = \frac{\hat{p}^2}{2m} + V(\hat{r})$, the symmetry is described by the direct product of the space group $G$ and the orthogonal group $SO(3) \times Z_2^T$ in spin space, reflecting the complete independence of spin and spatial degrees of freedom. Upon the introduction of SOC term $H_{soc} \propto (\nabla V(\hat{r}) \times \hat{p}) \cdot \hat{\sigma}$, the $SO(3)$ symmetry in spin space is broken, necessitating the use of non-magnetic double space groups $G \times Z_2^T$ to characterize the band representations. Furthermore, the introduction of magnetic order $H_{mag} = S(\hat{r}) \cdot \hat{\sigma}$ breaks time-reversal symmetry $T$. In this regime, the SOC term restricts symmetry operations to those that both spin and lattice rotate about the same axis by the same angle. This requirement means that every rotational operator acts on both the lattice and spin spaces, defining the traditional framework of MSGs. However, in the non-relativistic limit where SOC is negligible, spin and lattice spaces are partially decoupled, allowing a more comprehensive SSG framework to describe the underlying magnetic geometry (see Box 1). From this perspective, MSGs can be understood as subgroups of SSGs, describing symmetry-broken states induced by SOC.

SSGs establish a precise mapping between magnetic structures and corresponding symmetry groups, addressing the long-standing challenge in magnetic group theory[11]. While traditional MSGs cannot formally distinguish between the phenomenological categories of FM and AFM, SSGs provide a rigorous symmetry-based classification of magnetic orders[14]. Rooted in Néel's original conception[33], AFMs can be strictly defined as magnets where the net magnetic moment is constrained to zero by symmetry, otherwise belonging to FMs or ferrimagnets. Furthermore, the inclusion of spin translation groups[34], which combine spin rotations with lattice translations, enables the precise description of magnetic propagation modes. This allows for a systematic differentiation of complex magnetic orders, such as Néel-type, helical and multiple-$q$ magnets[14].

From the perspective of physical properties, SSGs naturally disentangle the effects induced by magnetic geometry from those induced by SOC[14,21-25]. This separation provides a powerful platform for engineering spin textures in reciprocal space and for manipulating quantum geometry directly from real-space magnetic geometry. On the other hand, MSGs, as subgroups of SSGs, describe the modulation of these physical properties when SOC is introduced. To bridge these two regimes, the recently developed oriented SSG framework[14], which aligns the coordinate between spin and lattice space, provides a unified symmetry language that captures both magnetic geometry and SOC effects.

This new paradigm has catalyzed the discovery of unconventional magnetism, a class of materials that host compensated magnetic structures yet manifest $T$-odd functionalities. These unconventional magnets offer a unique opportunity to merge the high-speed, high-density advantages of AFMs with the facile read-write capability of FMs. In the following sections, we introduce these ferromagnetic-like features from a symmetry perspective, focusing on both cases: with and without SOC. We elucidate how real-space magnetic geometry, through SSG symmetries, dictates the spin textures, quantum geometry and emergent quasiparticles, giving rise to the emergence of various $T$-odd ferromagnetic phenomena.

## III. Emergent functionalities of unconventional magnetism
### *Spin splitting and spin polarization*

As a classical $T$-odd signature, spin splitting presents the energy separation between the two spin channels in momentum space, expressed as $\Delta E(\bm{k}) = E(\bm{k},\uparrow) - E(\bm{k},\downarrow)$. In non-magnetic crystals without SOC, the spin-space $SO(3) \times Z_2^T$ symmetry at any arbitrary wavevectors enforces the spin degeneracy throughout the Brillouin zone. The emergence of collinear ferromagnetic order breaks $T$ and reduces the symmetry to $G \times SO(2) \rtimes Z_2^K$, permitting Zeeman-type splitting. This fundamental mechanism underpinned the giant magnetoresistance effect, marking the birth of conventional spintronics based on ferromagnets[35,36].

Considering SOC, the combination of spatial inversion $P$ and time reversal $T$ (i.e., $PT$ symmetry) enforces spin degeneracy throughout the Brillouin zone (Table I). Consequently, spin splitting in nonmagnetic materials with a general $G \times Z_2^T$ group manifests only if the system is non-centrosymmetric, allowing SOC to induce odd-parity textures (i.e., $\boldsymbol{p}(\boldsymbol{k},\uparrow) = -\boldsymbol{p}(\boldsymbol{k},\downarrow)$) such as Rashba or Dresselhaus splitting[37,38]. This structural asymmetry enables charge-to-spin conversion via spin-orbit torques[39,40], propelling spintronics into an era defined by the efficient manipulation of magnetic states via relativistic SOC effects.

SOC is not the only source of momentum-dependent spin splitting in non-magnetic systems; strong electronic correlations can trigger Pomeranchuk instabilities, leading to spontaneous symmetry breaking and the emergence of various spin textures[41,42]. Indeed, such correlation-induced spin textures represent a theoretical precursor to the spin-split unconventional magnets. However, these theoretical models lacked a direct connection to real-space magnetic geometries, hindering the identification of material candidates. An alternative route to non-relativistic spin splitting lies in AFMs with a compensated magnetization[43-48]. Initially, the MSG framework was employed to identify spin splitting in AFMs. However, while the MSG framework happens to be suited to collinear AFMs, it does not yield rigorous symmetry criteria for spin splitting in non-collinear magnetic configurations[22,25,49]. This practical limitation has necessitated the adoption of the SSG framework to systematically identify spin splitting induced by magnetic order.

One prototypical class of spin-split AFMs is the collinear type, which is now widely recognized as altermagnets[21,50-52]. In these systems, magnetic sublattices with opposite spins are connected by rotation/roto-inversion operations $A$ rather than $P$ or $\tau$. Their symmetry is described by the 10/37/422 collinear spin Laue[21]/point[21]/space[53] groups with the structure $(\{E\|H\} + \{T\|AH\}) \times SO(2) \rtimes Z_2^K$. Since $SO(2)$ symmetry preserves $\sigma_z$ as a good quantum number, the lifting of spin degeneracy requires the simultaneous breaking of both $PT$ and $U_n(\pi)\tau$. Furthermore, the $Z_2^K$ symmetry restricts these altermagnets to even-parity spin textures (i.e., $p(k,\uparrow) = p(k,\downarrow)$), such as $d$, $g$ or $i$-wave, as shown in Fig. 3a. Following the theoretical prediction of various material candidates[21,47], altermagnetic spin splitting has recently been observed via angle-resolved photoemission spectroscopy in $g$-wave altermagnets like MnTe and CrSb[54-59], as well as $d$-wave candidates including $KV_2Se_2O$ and $Rb_{1-\delta}V_2Te_2O$[60,61].

Spin-split electronic structures are not exclusive to collinear altermagnets; they are equally attainable in non-collinear AFMs, where the absence of the spin-space $SO(2)$ symmetry renders more complicated constraints. The first and most direct category consists of noncollinear AFMs that break both PT and $U_n\tau$. Representative materials include the $Mn_3X$ (X = Sn, Ir, Ge, Pt) family[28,62]. More recently, spin-polarized bands in non-coplanar AFM $MnTe_2$ have been predicted (Fig. 3b) and observed via spin- and angle-resolved photoemission spectroscopy[63]. Another class is characterized by the presence of a single $U_n\tau$ operation[22], which protects the degeneracy of the spin component perpendicular to the rotational axis $n$, while leaving the remaining spin component $p_n(k)$ free to split. This leads to a collinear spin texture aligned with the spin-rotation axis. In contrast to altermagnets, non-collinear spin-split states can exhibit either odd or even parity[64-66]. In coplanar spiral AFMs with $U_n\tau$ symmetry (Fig. 3c), $TU_n(\pi)$ enforces odd-parity collinear spin texture. This symmetry constraint remains valid regardless of the presence of $T\tau$, distinguishing from collinear altermagnets where $TU_n(\pi)$ dictates even-parity spin textures. For example, the odd-pairty spin texture is predicted in 120-degree spiral AFM $Ba_3MnNb_2O_9$[67], where $T\tau$ is absent. Recently, non-collinear AFMs with $U_n(\pi)\tau$ and $T\tau$ symmetry are referred to as $p$-wave magnets[49]. On the other hand, when multiple $U\tau$ operations constitute a $D_2$ group in spin space at any wavevector (Table I), spin degeneracy can be restored throughout the Brillouin[22], as exemplified by the non-coplanar AFMs $CoX_3S_6$ (X = Ta, Nb)[25,68,69] and strained $\gamma$-$Fe_xMn_{1-x}$[70].

The spin-split AFM can be extended to low-dimensional magnets; the reduced dimensionality of momentum space imposes additional symmetry constraints. For example, 2D altermagnets require the additional breaking of symmetries such as $\{T\|C_{2z}\}$ and $\{U_n(\pi)\|M_z\}$ because of the absence of the $k_z$ direction in momentum space[71]. These requirements significantly restrict the number of altermagnetic symmetry groups, resulting in 7/26/92 spin Laue[72]/point/layer groups. Beyond these, compensated ferrimagnets represent another subclass of unconventional magnets, characterized by zero net magnetization and ferromagnetic SSG symmetries. Due to the absence of symmetry operations connecting distinct magnetic sublattices, these materials exhibit Zeeman-type spin splitting throughout the Brillouin zone (Fig. 3d), resembling FMs[73-75].

Furthermore, the concept of hidden spin polarization offers new perspectives by shifting the focus from global to local symmetries[76]. In non-magnetic systems, even when $P$ is preserved, the breaking of local inversion symmetry can induce spin polarization within specific real-space sectors[77-80]. This concept becomes even more profound in AFMs, where stronger symmetry constraints give rise to diverse hidden states[81]. For instance, a $PT$ AFM can be decomposed into hidden ferromagnets[82,83] or hidden altermagnets[84] based on the local environment. Beyond breaking global $PT$ symmetry, spin polarization can also be achieved by lifting the local $U\tau$ symmetries[85].

The importance of spin polarization in unconventional magnets lies in the potential of electrical detection and manipulation of magnetic order, which are essential for information readout and writing. For example, detection can be realized through the tunneling magnetoresistance effect[86-90], where conductance is governed by the matching of spin-polarized Fermi surfaces between two electrodes in momentum space. The manipulation can be achieved through the generation of spin currents, such as the magnetic spin Hall effect and spin-splitting torque[30,32,91,92]. Unlike SOC-dominated phenomena, these effects originate from the magnetic geometry, thereby circumventing the trade-off between charge-spin conversion efficiency and spin diffusion length[21,93].

### *Quantum geometry*

Besides spin textures, magnetic geometry also dictates the quantum geometric properties of electronic wavefunctions, providing additional pivots for the electrical detection and manipulation of magnetic order. Following ref. [94], the quantum geometric tensor[94-98] for a Bloch eigenstate $|u_n(k)\rangle$ with band index $n$ is defined as

$$Q_n^{\alpha\beta}(k) = \langle \partial_{k_\alpha} u_n(k) | [1 - |u_n(k)\rangle\langle u_n(k)|] | \partial_{k_\beta} u_n(k) \rangle = g_n^{\alpha\beta}(k) - \frac{i}{2}\Omega_n^{\alpha\beta}(k),$$

where the imaginary part of $Q_n^{\alpha\beta}$ corresponds to the Berry curvature $\Omega_n^{\alpha\beta}$, and the real part gives the quantum metric $g_n^{\alpha\beta}$. Exerting $T$ operation gives rise to $\Omega_n^{\alpha\beta}(k) = -\Omega_n^{\alpha\beta}(-k)$ and $g_n^{\alpha\beta}(k) = g_n^{\alpha\beta}(-k)$, indicating that Berry curvature is $T$-odd whereas quantum metric is $T$-even. In addition, for the derivatives of quantum geometry, the Berry curvature dipole[99] $\nabla_k \Omega_n^{\alpha\beta}(k)$ becomes $T$-even, while the quantum metric dipole[100,101] (QMD) $\nabla_k g_n^{\alpha\beta}(k)$ is $T$-odd. The Berry curvature governs the intrinsic linear anomalous Hall effect[102] (AHE). Similarly, the band-renormalized QMD[103], also known as Berry curvature polarizability[104], underlies intrinsic nonlinear transport, providing a detectable signature for the identification of magnetic orders[103,104]. Recently, these $T$-odd quantum geometry effects have also been identified in AFMs, establishing them as a distinct facet of unconventional magnetism.

As the prototypical example of quantum geometric transport, we first focus on the Berry curvature induced AHE. While the AHE was traditionally regarded as the experimental hallmark of FMs[102,105], recent theoretical predictions and experimental observations have confirmed its presence in AFMs (Figs. 4a and 4b)[106-110]. In fact, the prerequisite for the AHE is not the existence of a net spin magnetization, but rather a net orbital magnetization allowed by symmetry. In the presence of SOC, a material is AHE-active if its magnetic point group permits a net magnetization—a condition satisfied by 31 of the 122 magnetic point groups, which we refer to as ferromagnetic magnetic point groups[111]. For example, in the coplanar AFM $Mn_3Sn$, even if its out-of-plane spin magnetization is assumed to vanish, its magnetic point group (*m'm'm*) is ferromagnetic, thereby permitting a net Berry curvature and magnetization along the *z*-direction. Such materials must inevitably exhibit SOC-induced weak FM, comprising both orbital and spin magnetization. However, the dependence of these two contributions on SOC may differ substantially[14].

In the absence of SOC, orbital and spin magnetizations become disentangled and are subject to distinct symmetry constraints, necessitating the SSG framework for a proper description[22,25]. The symmetries that permit a net orbital magnetization and, consequently, the AHE require mapping the SSG to an effective magnetic point group, which must be ferromagnetic. This mapping is defined by the correspondence $\{U\|R|\tau\} \to R$ and $\{TU\|R|\tau\} \to R'$, where the spin operation $U$ is absorbed by the proper lattice rotation $R$ as a magnetic point group operation. In collinear and coplanar magnetic structures, the spin-only symmetry $TU_n(\pi)$ acts as the effective $T$ for the orbital magnetization, thereby forbidding the AHE. As a result, SOC-free AHE induced purely by magnetic geometry can only emerge in noncoplanar magnets[22,25,69,106,112]. Such magnetic symmetries have also been referred to as scalar spin chirality[106,109]. A representative example is CoTa$_3$S$_6$, whose SSG $P^{3^2_{-1}-1}6_3{}^{m_{110}}2{}^{m_{011}}2|(2_{001},2_{100},1)$ in the Chen-Liu nomenclature[22], yields an effective magnetic point group $62'2'$, permitting a net orbital magnetization along the $z$-direction (Fig. 4c). A counterintuitive consequence is that, in the absence of SOC, a collective rotation of the local spin moments does not change the direction of the AHE. This highlights that the AHE arises entirely from the Berry curvature field generated by lattice symmetry and exchange interactions, rather than from SOC.

Although both are $T$-odd phenomena, the AHE, which is an integrated quantity over the Brillouin zone, and momentum-dependent spin splitting are governed by fundamentally different symmetry constraints (Table I), giving rise to distinct material pools within unconventional magnets[26]. Only in three-dimensional collinear and coplanar magnets do anomalous-Hall AFMs form a subset of spin-split AFMs. In non-coplanar and two-dimensional magnets, however, anomalous-Hall AFMs can exist without accompanying spin splitting[22,25,113]. Moreover, SOC-free spin splitting and SOC-induced magnetization—along with the associated AHE—may coexist in the same material. Recently, materials that exhibit non-zero orbital magnetization under SOC have been classified as spin–orbit magnets[14].

Next, we turn to nonlinear transport induced by higher-order quantum geometry effects. Within the framework of relaxation-time ($\tau_{rel}$) approximation, the extrinsic nonlinear Hall effect (proportional to $\tau_{rel}^1$) driven by $T$-even Berry curvature dipole was theoretically predicted[99] and then experimentally observed in non-magnetic WTe$_2$[114]. Subsequently, the intrinsic nonlinear response (proportional to $\tau_{rel}^0$) associated with the $T$-odd QMD[100] was predicted in AFMs such as CuMnAs, Mn$_2$Au and MnBi$_2$Te$_4$[103,104,115] (Fig. 4d), and has been experimentally demonstrated in AFM MnBi$_2$Te$_4$[101,116]. As shown in Fig 4e, in a four-layer MnBi$_2$Te$_4$ device, applying an alternating current at frequency $\omega$ produces a nonlinear Hall voltage at frequency $2\omega$ across the transverse direction[101,116]. When the Néel vector of the sample is reversed, this nonlinear Hall voltage also changes its sign, indicating that the antiferromagnetic order can be probed via QMD-based nonlinear transport.

While a nonzero QMD requires both *P* and *T* symmetries to be broken, intriguingly, it can survive under *PT* symmetry. This is in sharp contrast to both cases of spin splitting and the AHE, which are forbidden in *PT*-symmetric AFMs. Consequently, it is fair to say that *PT*-symmetric AFMs that exhibit *T*-odd nonlinear transport signatures, such as CuMnAs and $MnBi_2Te_4$, can be regarded as another distinct class of unconventional magnets in the context of quantum geometry. Indeed, the first experimental demonstration that a digital bit can be electrically written and read by manipulating and detecting the Néel order was reported in CuMnAs[9,117].

Similar to the linear AHE, *T*-odd nonlinear transport arising from quantum geometry typically results from the interplay between magnetic exchange and SOC. To obtain the purely magnetic-geometry contribution to quantum geometry in the absence of SOC, the SSG framework is required. Analogous to the Berry curvature, in collinear or coplanar AFMs such as the aforementioned $MnBi_2Te_4$, the spin-only symmetry $TU_n(\pi)$ serves as an effective *T*, forbidding the QMD. Therefore, SOC plays an indispensable role in any observed nonlinear transports in such systems. In noncoplanar magnetic materials, however, $TU_n(\pi)$ is broken, allowing the magnetic geometry alone to generate a finite QMD and drive nonlinear transport. Recent theoretical work has uncovered this magnetic-geometry-induced intrinsic nonlinear response through SSG-based symmetry analysis and first-principles calculations[118]. Notably, the resulting nonlinear conductivity can even be considerably larger than that involving SOC—for example, the calculated nonlinear conductivity of the noncoplanar AFM CrSe is 22.4 S²/A (Fig. 4f)[118], which is orders of magnitude larger than the ~0.01 S²/A observed in $MnBi_2Te_4$ thin films[116].

While divergent interpretations of the longitudinal component in QMD-induced nonlinear transport have emerged recently[119,120], it is worth noting that another mechanism, the Drude contribution ($\sim\tau_{rel}^2$), also constitutes a *T*-odd effect capable of probing antiferromagnetic order[82,117]. Beyond second-order transport, the third-order nonlinear current associated with the quantum geometry quadrupole has also recently become the subject of both theoretical and experimental investigation[121-125].

*Topology and quasiparticles*

The magnetic geometry of unconventional magnets dictates a rich landscape of topological states, which also provide signatures for the detection of magnetic orders. While long-range magnetic order inevitably breaks $T$, the $Z_2$ topological phase has been generalized to AFMs protected by $T\tau$ symmetry[126]. Representative examples include collinear AFMs such as $MnBi_2Te_4$ and $EuSn_2As_2$[127,128]. Furthermore, the theories of topological quantum chemistry and symmetry indicator under MSGs have been systematically developed and applied to high-throughput searches for AFM topological phases[129-131].

Compared with MSGs, SSGs provides more symmetries that can protect topological phases. For collinear and coplanar magnetic structures, there exists the spin-only operation $TU_n(\pi)$ that squares to one, which constrains the system to class AI of the Altland-Zirnbauer tenfold classification[23,53]. As a result, both Chern insulators and $Z_2$ topological insulators are forbidden. In contrast, the absence of spin-only operations in non-coplanar magnets permits the possibility of the quantum anomalous Hall effect even without SOC, as exemplified by $K_{0.5}RhO_2$[112]. Furthermore, SSGs support an expanded set of operations with squares of $\pm 1$ that are forbidden in MSGs, leading to more topological states beyond MSGs[19]. A representative example is the $Z_2$ magnetic topological insulator (Fig. 5a), which is protected by the $\{T\|PC_{2z}\}$ operation[19,23]. These band topologies arising from non-coplanar magnetic geometry provide additional degrees of freedom for the electrical detection and manipulation of magnetic orders.

Antiferromagnetic topological materials extend beyond insulators to semimetals characterized by Weyl or Dirac points near the Fermi level[132,133]. The Weyl semimetal was initially proposed in non-coplanar all-in-all-out antiferromagnetic iridates[134]. The Weyl point is defined by linear band crossings that carry a non-zero Chern number, and exhibiting singular monopoles of Berry curvature and topologically protected Fermi arcs. The Chern number can be expressed as the Berry curvature flux through a closed surface $S$ enclosing the Weyl point, $C = \frac{1}{2\pi}\oint_S \Omega(k) \cdot dS$. Consequently, Weyl points in AFMs require the breaking of both $PT$ and $PT\tau$ symmetries. The first experimental evidence of Weyl fermions in AFMs was reported in Mn$_3$Sn[62,135], where the Berry curvature at the Fermi level serves as the microscopic origin of its large anomalous Hall effect[108]. In contrast, Dirac semimetals can be viewed as a superposition of Weyl nodes with opposite Chern numbers[136,137], which requires the preserved $PT$ or $PT\tau$ symmetry to protect the band degeneracy throughout the whole Brillouin zone. In general, Dirac points lack net topological charges and symmetry-protected surface states.

The inclusion of spin group symmetries further enriches the landscape of topological quasiparticles. For instance, the SO(2) symmetry in altermagnets enables topological features characterized by a single spin, including spin-polarized Weyl points, nodal lines and nodal planes (Fig. 5b)[138,139]. Furthermore, the $U_n(\pi)\tau$ symmetry can protect the overall band degeneracy even in non-centrosymmetric collinear AFMs. Compared to $PT$ or $PT\tau$ symmetry protected Dirac points, the $U_n(\pi)\tau$ symmetry connects Weyl points with identical Chern numbers, leading to chiral Dirac fermions[140,141] with a net topological charge of two. The resultant doubly degenerate Fermi arcs are topologically protected, robust against perturbations.

More generally, SSGs not only accommodate larger topological charges but also higher band degeneracy than MSGs, including twelve-fold nodal points (Fig. 5c), eight-fold nodal lines, and four-fold nodal planes[53,142,143]. Such features are absent in the framework of MSG and will be further discussed in the context of topological magnons. Within the electronic systems, based on the irreducible co-representation theory of SSGs, 218 classes of quasiparticles have been identified[142]. Recently, symmetry indicator theory and symmetry-breaking analysis based on SSGs have been employed in the high-throughput search for topological states in magnetic materials[144]. In addition to exact symmetries, the quasi-symmetry group[145,146] has been introduced to SSGs, leading to the spin-orbit U(1) protected topological charge quadrupole[147]. This unconventional quasiparticle significantly enhances the nonlinear Hall response, providing a new mechanism for detecting and manipulating antiferromagnetic order.

SSG symmetries further extend the topological phases to bosonic excitations, such as topological magnons. Although topological magnons have been widely predicted[148,149], their experimental realization in AFMs remains limited[150-152]. Take the *PT*-symmetric AFM $Cu_3TeO_6$ as an example: Dzyaloshinskii-Moriya interactions are predicted to induce Weyl nodal lines and non-degenerate band at general wavevectors[153]. However, six-fold magnons, Dirac magnons at certain wavevectors, and doubly degeneracy throughout the Brillouin zone have been observed by inelastic neutron scattering[154,155]. This discrepancy indicates that magnon band structures in Heisenberg-dominated magnets are dictated by SSGs rather than MSGs[16,53]. Indeed, the concept of SSG was first proposed by Brinkmann and Elliott in 1966[16], following their observation of extra band degeneracies in magnon band structure that cannot be explained by the representation theory of MSGs.

Recent developments in band representation theory of SSGs have facilitated the exploration of unconventional magnons beyond MSGs[53]. These unconventional magnons encompass chiral magnons[156] along with nodal points, lines and planes characterized by higher topological charge or higher dimensionalities. For example, in collinear AFMs, the $U_n(\pi)\tau$ symmetry ensures that the opposite spin channels possess identical topological charges, yielding charge-8 Dirac magnons and charge-4 octuple magnons. Systematic diagnosis based on SSG symmetries has identified 500 material candidates hosting unconventional magnons and established a comprehensive magnon band database (Fig. 5d). Recently, chiral magnons have been observed in several altermagnets by inelastic neutron scattering and spin transport measurements[157-161].

The influence of SOC on magnon band structure is introduced through SOC-related exchange interactions. These terms do not directly break the SSG into the MSG but instead induce richer symmetry-breaking paths. For instance, the inclusion of Kitaev interactions in a Heisenberg model partially contrains the SO(2) symmetry to two-fold spin rotation[20]. Consequently, while the double degeneracy throughout the Brillouin zone is lifted and the four-fold nodal line is evolved into four-fold nodal points, such band features remain beyond the description of MSGs (Fig. 5e). Similarly, in two-dimensional magnets, out-of-plane Dzyaloshinskii-Moriya interactions can break the $TU_n(\pi)$ operation while preserving SO(2) symmetry[162], facilitating the realization of pure magnon spin currents.

**IV. Outlook**

This review provides an overview of unconventional magnetism, defined by the combination of compensated magnetization and *T*-odd physical responses. By employing SSG theory, we elucidate how magnetic geometry dictates three pivotal facets: spin textures, quantum geometry, and emergent quasiparticles, thereby illustrating the potential of unconventional magnets to enrich the landscape of spintronics. In the following, we discuss the intriguing directions that emerge when unconventional magnetism is coupled with other material attributes, such as ferroelectricity, superconductivity, and moiré systems.

The deterministic switching of *T*-odd physical responses is fundamental to spintronic applications, where electrical control is particularly favored for its high integration density and energy efficiency. While the manipulation of AFM order has been demonstrated using different spin torques[8,9,90,163-165], establishing a comprehensive methodology to govern the electrical switching in unconventional magnets remains an open challenge. Current symmetry analyses indicate that the specific tensor forms of spin torques are dictated by the operations connecting magnetic sublattices[166], while deterministic switching further necessitates breaking inversion symmetry to lift path degeneracy[167]. On the other hand, analogous to FMs, the *T*-odd signatures of unconventional magnets facilitate their integration with additional ferroic orders, such as ferroelectricity and ferroelasticity, providing versatile methods to probe and manipulate magnetic order. For example, coupling ferroelectricity with spin-splitting AFMs enables the simultaneous switching of electric and spin polarization[168,169], which has been recently demonstrated[170]. Furthermore, modulating lattice symmetry through ferroelastic strain or sliding mechanisms can trigger phase transitions between distinct magnetic orders, enabling the non-volatile switching of multifunctional states[171-173]. The cooperation among these multiple degrees of freedom not only elucidates the interplay of symmetry-dictated properties but also suggests a route toward all-electrical, low-power topological spintronic architectures.

Unconventional magnetism also offers a unique platform to explore quantum coherent phenomena through its interplay with superconductivity. The spin-momentum locking in unconventional magnets provides the opportunity for the realization of many exotic superconducting states, such as the unconventional Andreev reflection[174,175], anomalous Josephson effects[176,177], the superconducting diode effect[178] and the topological superconductivity[179]. Furthermore, the coupling of magnetic geometry with superconductors allows for the generation of spin-polarized supercurrents and triplet pairing[180,181], alongside the finite-momentum pairing even in the absence of net magnetization[182]. Such mechanisms not only enrich the landscape of unconventional superconductors but also offer strategies to enhance critical superconducting parameters[183]. While theoretical models have mapped out an extensive landscape of these quantum phenomena, bridging the gap between theoretical predictions and experimental demonstration remains the primary challenge for implementing unconventional magnetism in future superconducting technologies.

The landscape of unconventional magnetism with tailored $T$-odd responses can be further extended to moiré interfaces and amorphous systems beyond crystal symmetry. Twisted van der Waals bilayers serve as a universal platform for inducing unconventional magnets with conventional FMs or AFMs[184,185]. This approach also enables the coupling of magnetic order with sliding/ moiré ferroelectricity[186] and valley degrees of freedom[187]. Such phenomena are not restricted to periodic lattices but extend to quasicrystals, where the exploitation of higher non-crystalline rotational symmetries protects exotic anisotropic spin-splitting states[188-190].

Looking forward, unconventional magnetism is a nascent field with vast potential for exploration. The integration of spin space group theory with first-principles-based high-throughput screening[191-193], augmented by machine-learning methods[194,195], provides a pathway for the systematic discovery of unconventional magnets. These computational strategies aim to accelerate the identification of unconventional magnets that possess substantial $T$-odd responses and high Néel temperatures, thereby facilitating the development of robust writing and reading components for spintronic devices. Furthermore, investigating responses under extreme conditions or at ultrafast timescales may unveil novel physical mechanisms and dynamic control protocols. Such targeted endeavors are essential for transitioning from fundamental explorations to the practical realization of spintronics and quantum information technologies.

**Acknowledgements**

We thank P. Liu, Y. Liu, J. Li and T. Zou for the helpful discussions. This work was supported by National Natural Science Foundation of China under Grant No. 12525410, No. 12274194, No. 12574275 and No. 12534003, Guangdong Provincial Quantum Science Strategic Initiative under Grant No. GDZX2401002, Guangdong Provincial Key Laboratory for Computational Science and Material Design under Grant No. 2019B030301001, Shenzhen Science and Technology Program (Grant No. RCJC20221008092722009 and No. 20231117091158001), the Innovative Team of General Higher Educational Institutes in Guangdong Province (Grant No. 2020KCXTD001) and Center for Computational Science and Engineering of Southern University of Science and Technology.

**Box 1 | Basics of Spin space group.**

In the absence of SOC, the full symmetry group (considering spin) of a non-magnetic system has the direct product form $G \times SO(3) \times Z_2^T$, where $G$ stands for the space group in lattice space, $SO(3)$ is the full rotation group in spin space, and the binary group $Z_2^T = \{E, T\}$ contains time reversal $T$. An SSG is defined as a subgroup of $G \times SO(3) \times Z_2^T$ that leaves a magnetic structure invariant. A group element of an SSG is represented by $\{g_S \| g_R | \tau\}$, where $g_S$ and $g_R$ represent proper/improper rotations in spin and lattice spaces, respectively, and $\tau$ corresponds to the fractional translation in lattice space. Note that improper rotations are expressed as the combination of proper rotations with $T$ in spin space or spatial inversion $P$ in lattice space.

An SSG can also be written in a direct product form between the non-trivial SSG $G_{NSS}$ and the spin-only group $G_{SO}$, i.e., $G_{NSS} \times G_{SO}$. $G_{NSS}$ contains every spin operation combined with a spatial operation (except for the identity), while $G_{SO}$ consists of pure spin operations, characterizing the dimensionality of the magnetic geometry. For collinear magnets, $G_{SO}^l = SO(2) \rtimes Z_2^K$, where $SO(2)$ stands for the continuous rotation group around the collinear axis, $Z_2^K = \{E, TU_n(\pi)\}$ contains the spin mirror symmetry $TU_n(\pi) = K$ (**n** is any direction perpendicular to the collinear axis, $K$ denotes complex conjugation). For coplanar and non-coplanar magnets, the spin-only group are restricted to $G_{SO}^p = Z_2^K$ (**n** is the direction parallel to the normal line of the spin plane) and $G_{SO}^n = \{E\}$, respectively.

**Table I**. Symmetry criteria for momentum-dependent spin splitting and the anomalous Hall effect in AFMs. Effective-MPG: The effective magnetic point group (MPG) derived by mapping SSG elements according to $\{U\|R|\tau\} \to R$ and $\{TU\|R|\tau\} \to TR$. $U_n(\pi)\tau$: the combination operation of two-fold spin rotation and fractional translation. *PT*: combination of space inversion and time reversal, here *PT* includes $PT\tau$ ($\tau$ stands for fractional translation).

| AFM | Spin splitting | | Anomalous Hall effect | |
|---|---|---|---|---|
| | w/o SOC | w/ SOC | w/o SOC | w/ SOC |
| Collinear | break *PT* and $U_n(\pi)\tau$ | break *PT* | disallowed | Ferromagnetic MPG |
| Coplanar | break *PT* and multiple $U_n(\pi)\tau$ | | | |
| Noncoplanar | | | Ferromagnetic effective-MPG | |

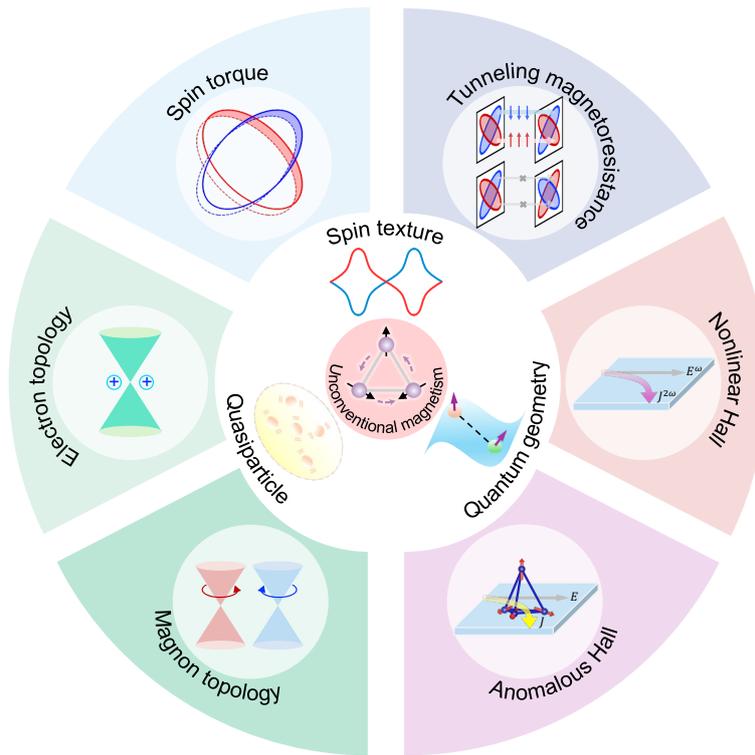

**Figure 1 | Schematic landscape of unconventional magnetism and its emergent phenomena.** The central schematics highlight unconventional magnetism, where the underlying magnetic geometry is characterized by SSGs. The inner ring represents the three fundamental facets: spin texture, quantum geometry, and emergent quasiparticle. These core elements give rise to the diverse $T$-odd emergent phenomena shown in the outer ring.

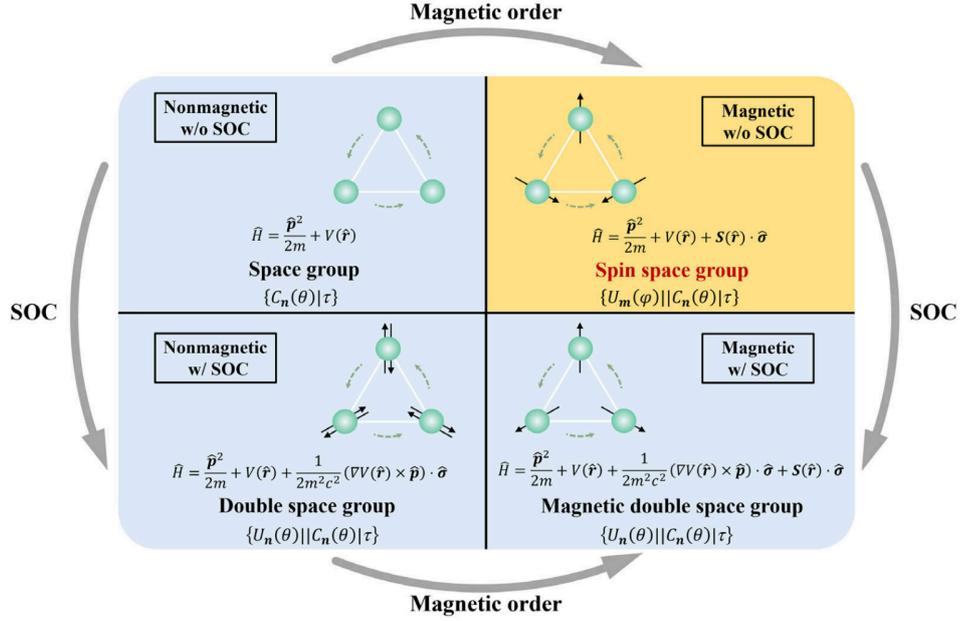

**Figure 2 | Four-quadrant diagram describing the symmetry of solids with/without magnetic order and/or SOC.** The general steady-state Hamiltonians, space groups, and their representative group elements are shown for each quadrant. Compared with the conventional crystallographic groups, the key characteristic of spin group is the partial decoupling between spatial rotation $C_n(\theta)$ and spin rotation $U_m(\varphi)$, where $\boldsymbol{m}$ and $\boldsymbol{n}$ denote the rotation axes, and the real scalars $\varphi$ and $\theta$ are the rotation angles. For the materials with SOC, spatial and spin rotations are completely locked. For the materials without SOC, the spin and spatial rotations are completely or partially decoupled, which implies that one symmetry operation could be composed of a spin and a spatial rotation with different rotation axes and angles $\{U_m(\varphi)\|C_n(\theta)|\tau\}$. Reproduced from ref. [19].

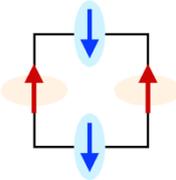

**Figure 3 | Symmetry-dictated spin splitting across different magnetic geometries.** **a.** Altermagnetic spin splitting in collinear AFMs with broken $PT$ and $U_n(\pi)\tau$ symmetries, yielding even-parity, collinear spin textures. **b.** Antiferromagnetic spin splitting in non-collinear AFMs with broken $PT$ and $U_n\tau$ symmetries, facilitating either even- or odd-parity non-collinear spin textures. **c.** Antiferromagnetic spin splitting in non-collinear AFMs with broken $PT$ but preserved $U_n\tau$ symmetries, enabling either even- or odd-parity persistent (collinear) spin textures. **d.** Zeeman-type spin splitting in compensated ferrimagnets, where the absence of symmetry operations connecting distinct magnetic sublattices allows for spin-polarized bands despite vanishing net magnetization. Representative material candidates for each category are listed accordingly.

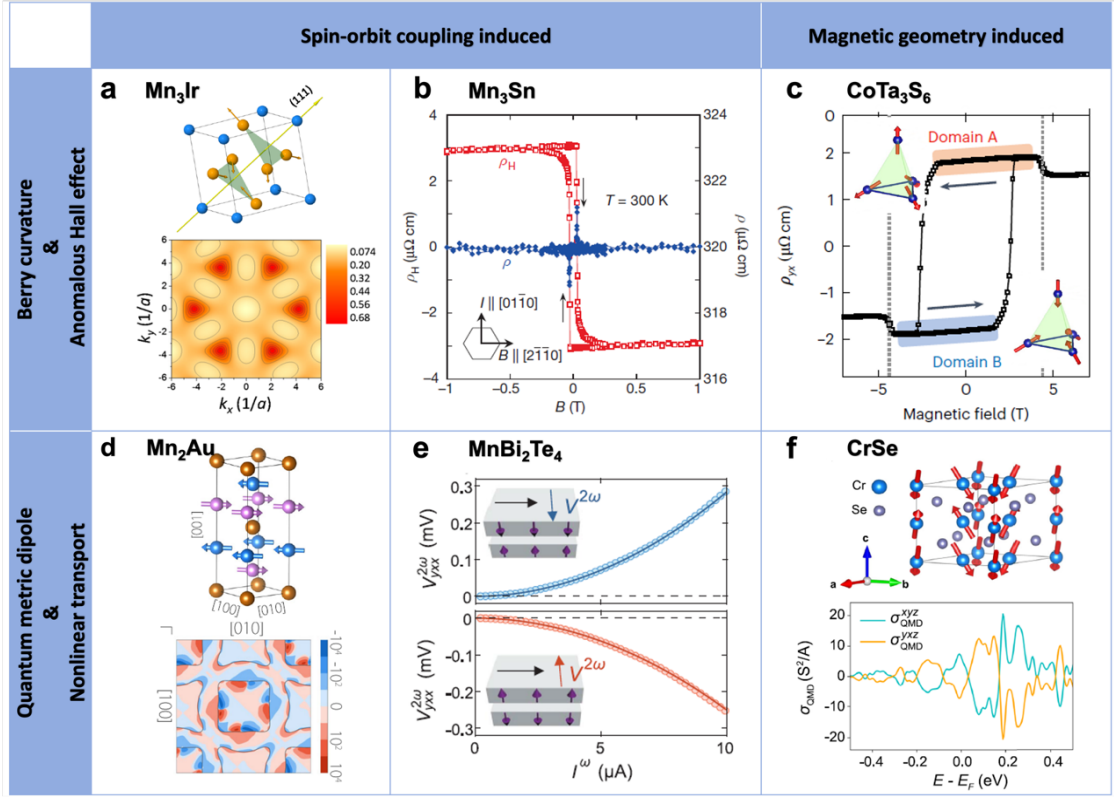

**Figure 4 | Quantum geometry and derived intrinsic quantum transport. a-c**, Berry curvature and anomalous Hall effect. **a**, Coplanar magnetic structure (left) and Berry curvature (right) of $Mn_3Ir$. Reproduced from ref. [107]. **b,c**, Measured hysteresis loops of anomalous Hall conductivity in coplanar $Mn_3Sn$ (b) and non-coplanar $CoTa_3S_6$. Reproduced from ref. [108] and [69]. (c), originating mainly from SOC and magnetic geometry, respectively. **d–f**, Quantum metric dipole and nonlinear transport. **d**, Collinear magnetic structure (left) and quantum metric dipole (right) of $PT$-symmetric $Mn_2Au$. Reproduced from ref. [104]. **e**, Measured nonlinear Hall conductivity of $MnBi_2Te_4$, with opposite signs for opposite Néel orders, induced by SOC. Reproduced from ref. [101]. **f**, Non-coplanar magnetic structure (top) and calculated nonlinear conductivity (bottom) of CrSe, driven by magnetic geometry. Reproduced from ref. [118].

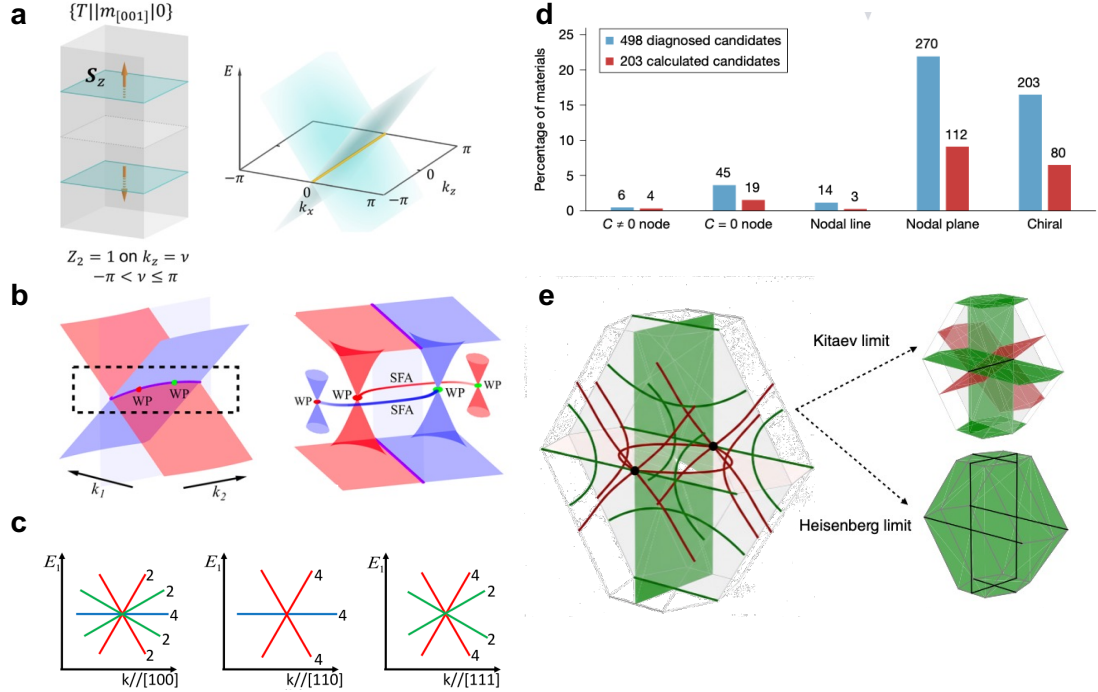

**Figure 5 | Symmetry-protected electronic and magnonic topology. a,** $Z_2$ magnetic topological insulators protected by $\{T\|m_{001}|0\}$ symmetry. Reproduced from ref. [19]. **b,** Opposite-spin (left) and same-spin (right) Weyl points. Reproduced from ref. [138]. **c,** The schematic dispersion of the 12-fold fermion along three high-symmetry lines [100], [110], [111]. Reproduced from ref. [142]. **d,** Statistics of five types of unconventional magnon in the collinear magnets including: (1) chiral quasiparticles (2) topological charge-neutral (C = 0) quasiparticles (3) octuple nodal line magnons; (4) quadruple nodal plane magnons and (5) chiral magnons. Reproduced from ref. [53]. **e,** Magnon band degeneracies within general Heisenberg-Kitaev model (left), Kitaev limit (upper right) and Heisenberg limit (lower right). Reproduced from ref. [20].

33